# Oscillatory tilt effect in a metal in a weak magnetic field

Oleg P. Ledenyov

*National Scientific Centre Kharkov Institute of Physics and Technology, Academicheskaya 1, Kharkov 61108, Ukraine.*

The oscillatory tilt effect in a normal metal at external magnetic field is discovered. The oscillatory tilt effect is characterized by the oscillations of ultrasound attenuation in a metal at external magnetic field as predicted in [1]. The dimension of the non-central cross-section of the *Fermi* surface and the velocity of electrons in a high pure *Gallium* single crystal at external magnetic field are found. At the low frequencies of ultrasonic signal, the "inverse" oscillatory tilt effect in a high pure *Gallium* single crystal at the external magnetic field is observed.

## Introduction

The tilt effect (*TE*) in the interaction of the conduction electrons with the ultrasound signal in a metal was discovered a long time ago [2], and it has been observed in a number of metals, including the *Gallium* [3]. The tilt effect appears as a sharp dependence of the ultrasound attenuation and of the ultrasound velocity dispersion on the angle $\varphi$ by which the magnetic field $H$ deviates from the direction orthogonal to the ultrasound propagation direction $q$. The tilt effect is referred as an effective method for the measurement of conduction electrons velocity at the *Fermi* surface, and has usually been researched in a metal at strong external magnetic fields at the condition $qR \ll 1$, where $q$ is the wave vector of ultrasonic wave, $\lambda$ is the ultrasonic wavelength, $|q|=2\pi/\lambda$, and $R$ is the *Larmor* radius of electron trajectory in a metal at external magnetic field.

The thorough theoretical researches on a possible existence of the tilt effect at weak magnetic fields at the condition $qR \gg 1$ were conducted in [1, 4, 5]. The new oscillatory tilt effect (*OTE*) in a metal at weak magnetic fields was predicted by Eremenko, Kaner, Fal'ko in [1]. The oscillatory tilt effect is an interesting subject of research, because it would make it possible to measure the dimensions of the non-central cross-sections of the *Fermi* surface in a metal in addition to the electron velocity measurement. The theory of the *OTE* in the metals and semi-metals with the multiply connected Fermi surfaces was further developed by Fal'ko, Kaner, Eremenko in [5]. It makes sense to note that, in the experimental part of research in [4], the oscillatory tilt effect in *Tungsten* at the external magnetic field of a few $kOe$ at the condition $qR > 1$ was not observed. The aim of this research was to try to observe the oscillatory tilt effect in a metal at weaker external magnetic field.

## Oscillatory tilt effect in a metal in a weak magnetic field

In the present paper, the research results on the dependence of the coefficient of ultrasound attenuation $\Gamma$ on the magnetic field deviation angle $\Gamma(\varphi)$ in the high pure *Ga* single crystal with the electron mean free path $l \sim 1\ cm$ is reported. The big magnitude of electron mean free path in the high pure *Ga* single crystal allowed to decrease the magnitude of external magnetic field $H$ by the two orders of magnitude, comparing to the magnitude of magnetic field $H$ in [4]. The precise ultrasonic measurements were carried out on the high pure *Ga* single crystal at the magnitude of external magnetic field $H = 42.3\ Oe$ at the temperature of *1.2 K*, using the calibrated measurement set up and applying the pulsed method. The cylindrical sample with the diameter of $\sim 1\ cm$ and length of about *2 cm* was cut from the high pure *Ga* single crystal of large dimensions by the electric erosion method. The acoustic transducers had the fundamental frequency of *10 MHz*, and were made of *X*-cut quartz and mounted on the polished flat end surfaces of a cylindrical sample with the help of the silicon-organic liquid. The logitudinal ultrasonic wave propagated along the axe *c* of a sample. The crystallographic orientation of a sample is adjusted within $\sim 0.1\,°$ by the *x*-ray techniques. In the measurement set up, the geomagnetic field is cancelled by the special magnetic system. The external magnetic field is created by a pair of the *Helmholtz* coils, which can be rotated in the plane perpendicular to the axis of a sample; it can also deviate from this plane by an angle up to $\varphi \sim 6\,°$. The research on the precise characterization of ultrasound propagation in a sample can be done at the big values of magnetic field deviation angle $\varphi$ due to the application of an auxiliary pair of the *Helmholtz* coils.

In the course of the experimental research, it was found that there is the oscillatory tilt effect in the high pure *Ga* single crystal at the weak external magnetic field at the low temperature. This oscillatory tilt effect is characterized by an oscillation in the dependence of the coefficient of ultrasound attenuation $\Gamma(\varphi)$ on the magnetic field deviation angle at small values of $\varphi$. The Fig. 1 shows the dependence of the coefficient of ultrasound attenuation on the magnetic field deviation angle $\Gamma(\varphi)$ in *Ga* single crystal. The frequency of longitudinal ultrasonic wave is $\omega/2\pi = 150$ MHz. The external magnetic field ***H*** is oriented along the axe ***a*** of a sample at the magnetic field deviation angle $\varphi = 0$; the external magnetic field ***H*** deviates in the direction of the axe ***c*** of a sample as the magnetic field deviation angle $\varphi$ is increased. The sharp peak, which is followed by the oscillations, in the dependence of the coefficient of ultrasound attenuation on the magnetic field deviation angle $\Gamma(\varphi)$ is clearly visible. The experimental results are in a good qualitative agreement with the theoretical prediction [1]. The value of $qR$ in the researched case is equal to *40*, and this value defines the sharpness of the first ultrasound attenuation peak at $\varphi = \varphi_k$. Using the relation $\sin\varphi_k = \dfrac{S}{v}$ in [1], it is possible to find the electron velocity $v \approx 4 \cdot 10^7 cm/sec$. This electron velocity agrees well with the value of electron velocity $v \approx 5 \cdot 10^7$ cm/sec, obtained from the measurements of the tilt effect in the *Ga* single crystal at the strong external magnetic fields in [3]. The periodicity of the oscillatory tilt effect is described by the expression in [1]

$$\Gamma_{oscil.} \sim \sin^2\left[qR\left(1 - \dfrac{\sin\varphi_k}{\sin\varphi}\right)^{1/2} - \dfrac{\pi}{4}\right].$$

Thus, it is possible to find the dimension of oscillating non-central cross-section of the *Fermi* surface, which is $\dfrac{p_b}{\hbar} = 1.4 \cdot 10^7 cm^{-1}$. The displacement of the oscillating non-central cross-section of the *Fermi* surface toward the direction of the axis ***a*** is $\dfrac{p_a}{\hbar} = 1 \cdot 10^7 cm^{-1}$, where $p_a$ and $p_b$ are the electron quasi-momentum on the *Fermi* surface along the directions ***a*** and ***b***. At the big magnetic field deviation angles $\varphi$, it was possible to observe the oscillations of the *Pippard* resonance, which begin to appear at the magnetic field deviation angles $\varphi \gtrsim 9°$ in an agreement with the calculations, using the expressions in [5] with the particular values of the *Fermi* surface in the *Ga* single crystal. This fact can be regarded as a clear evidence that the only oscillations generated by the oscillatory tilt effect can appear in the pure *Ga* single crystal at the small magnetic field deviation angles $\varphi$ as shown in Fig. 1.

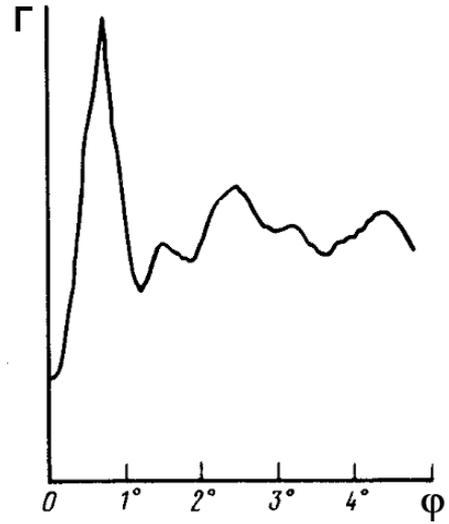

*Fig. 1. Dependence of ultrasound attenuation on magnetic field deviation angle $\varphi$ in high pure Ga single crystall* at temperature $T = 1.2$ K: $\omega/2\pi = 150$ MHz; ***q***//***c***; ***H***//***a*** at $\varphi = 0$.

In Fig. 2, the orientation dependence of the coefficient of ultrasound attenuation $\Gamma$ at the ultrasonic signal frequencies in the range of *30–50 MHz* at the rotation of the external magnetic field ***H*** in the plane, defined by the axes ***a*** and ***b***, (i.e., $\varphi = 0$) is observed.

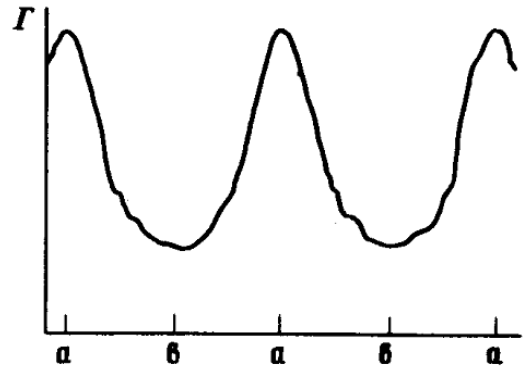

*Fig. 2. Dependence of coefficient of ultrasound attenuation $\Gamma$ on direction of magnetic field **H** in plane of axes **a**, **b** in high pure Gallium single crystal* at temperature $T = 1.2$ K; $\omega/2\pi = 50 MHz$; ***q***//***c***.

The one feature, which has to be highlighted, is an "inverse" tilt effect. It is observed, when the orientation of magnetic field is parallel to the direction ***H***||***a***, i.e. when the magnitude of the coefficient of ultrasound attenuation $\Gamma$ reaches its maximum. Fig. 3 shows the dependence of coefficient of ultrasound attenuation on deviation angle $\Gamma(\varphi)$ in high pure *Ga* single crystal at ***H***||***a*** at $\varphi = 0$ at $T = 1.2$ K. It is possible to notice that the coefficient of ultrasound attenuation changes in a manner opposite that in Fig. 1. The nature of differences in the dependences $\Gamma(\varphi)$ can be connected with the different physical mechanisms of ultrasound attenuation, assuming a presence of cyclotron resonance at *H*∥*a*, and a disappearence of cyclotron resonce as the magnetic field orientation *H* is changed. This effect can mask the ultrasound attenuation, connected with the oscillatory tilt effect or the tilt effect, and it appears in the type of dendence in Fig. 3. The "inverse" tilt effect requires a special research to be done, and the provided comment on its physical mechanism is simply tentative.

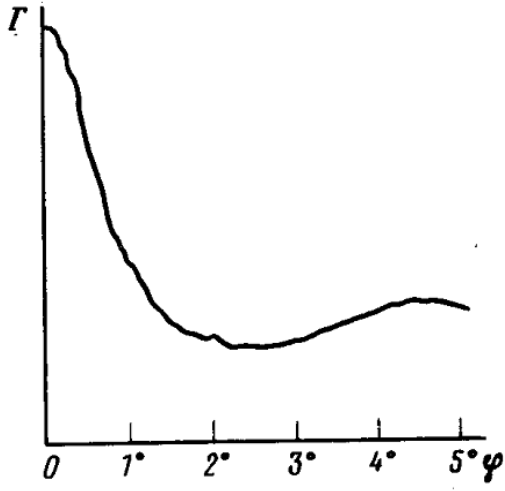

Fig. 3. *Dependence of coefficient of ultrasound attenuation on magnetic field deviation angle Γ(φ) in high pure Ga single crystal* at temperature *T = 1.2 K*:

$$\frac{\omega}{2\pi} = 30 \; MHz \; ; \; q \, \| \, c; \; H \, \| \, a \; at \; \varphi = 0.$$

## Conclusion

The oscillatory tilt effect in a normal metal at the external magnetic field *H* is discovered. The oscillatory tilt effect is characterized by the oscillations of the coefficient of ultrasound attenuation in a metal at the external magnetic *H* field as predicted in [1]. The dimension of the non-central cross-section of the *Fermi* surface and the velocity of electrons in a high pure *Ga* single crystal at the external magnetic field *H* are found. At the low frequencies of ultrasonic signal, the "inverse" oscillatory tilt effect in a high pure *Ga* single crystal at the external magnetic field *H* is observed.

This research paper was published in the *Letters to Journal of Experimental and Theoretical Physics (JETP Letters)* in 1986 [7].

[*]E-mail: ledenyov@kipt.kharkov.ua

———————